\documentclass[manuscript]{aastex}

\usepackage{natbib}
\usepackage{xcolor}

\renewcommand{\footnotesize}{\scriptsize}

\shorttitle{A new binary quasar}
\shortauthors{Altamura et al.}

\begin{document}


\title{Serendipitous discovery of a physical binary quasar at $z=1.76$}

\author{E. Altamura\altaffilmark{1}}
\affil{Jodrell Bank Centre for Astrophysics, Department of Physics and Astronomy,\\ The University of Manchester, Oxford Road, Manchester M13 9PL, United Kingdom}

\author{S. Brennan\altaffilmark{2}}
\affil{School of Physics, O’Brien Centre for Science North, University College Dublin, Belfield, Dublin 4, Ireland}

\author{A. Le{\'s}niewska\altaffilmark{3}}
\affil{Astronomical Observatory Institute, Faculty of Physics, Adam Mickiewicz University, ul. S{\l}oneczna 36, Pozna{\'n}, Poland
}

\author{V. Pint{\'e}r\altaffilmark{4}}
\affil{Doctoral School of Sciences, The University of Craiova, Str. A. I. Cuza nr. 13, 200585 Craiova, Romania
}

\author{S. N. dos Reis{\altaffilmark{5}}
\affil{
Departamento de F{\'i}sica, Faculdade de Ci{\^e}ncias da Universidade de Lisboa, Edif{\'i}cio C8, Campo Grande, 1749-016 Lisboa, Portugal
\and
Instituto de Astrof{\'i}sica e Ci{\^e}ncias do Espa\c{c}o - Observat{\'o}rio Astron{\'o}mico de Lisboa, Tapada da Ajuda, 1349-018 Lisboa, Portugal
}
}

\author{T. Pursimo{\altaffilmark{6}}
\affil{Nordic Optical Telescope, Apartado 474, E-38700 Santa Cruz de La Palma, Spain}
}

\author{J.~P.~U.~Fynbo\altaffilmark{7,8}}
\affil{
Cosmic DAWN Center NBI/DTU-Space
\and
Niels Bohr Institute, University of Copenhagen, Lyngbyvej 2, 2100 Copenhagen \O, Denmark
}

\author{S.~Geier\altaffilmark{9,10}}
\affil{Instituto de Astrofisica de Canarias, C/ Via Lactea, s/n, 38205, La Laguna, Tenerife, Spain 
\and
Gran Telescopio Canarias (GRANTECAN), 38205 San Cristóbal de La Laguna, Tenerife, Spain}

\author{K.~E.~Heintz\altaffilmark{11}}
\affil{Centre for Astrophysics and Cosmology, Science Institute, University of Iceland, Dunhagi 5, 107 Reykjav\'ik, Iceland}

\author{P.~M\o ller\altaffilmark{12}}
\affil{European Southern Observatory, Karl-Schwarzschildstrasse 2, D-85748 Garching, Germany}


\begin{abstract}
\noindent 
Binary quasars are extremely rare objects, used to investigate clustering on very small scales at different redshifts. The cases where the two quasar components are gravitationally bound, known as physical binary quasars, can also exhibit enhanced astrophysical activity and therefore are of particular scientific interest. Here we present the serendipitous discovery of a physical pair of quasars with an angular separation of $\Delta\theta = (8.76 \pm 0.11)$ arcsec. The redshifts of the two quasars are consistent within the errors and measured as $z = (1.76\pm 0.01)$. Under the motivated assumption that the pair does not arise from a single gravitationally lensed quasar, the resulting projected physical separation was estimated as $(76 \pm 1)$ kpc. For both targets we detected \ion{Si}{4}, \ion{C}{4},  \ion{C}{3]}, and \ion{Mg}{2} emission lines. However, the two quasars show significantly different optical colours, one being among the most reddened quasars at $z > 1.5$ and the other with colours consistent with typical quasar colours at the same redshift. Therefore it is ruled out that the sources are a lensed system. This is our second serendipitous discovery of a pair of two quasars with different colours, having a separation $\lesssim 10$ arcsec, which extends the very limited catalogue of known quasar pairs. We ultimately argue that the number of binary quasars may have been significantly underestimated in previous photometric surveys, due to the bias arising from paired quasars with very different colours.
\end{abstract}





\section{Introduction} \label{sec:introduction}


Over the past decade, several studies focused on characterising and selecting quasars, using both optical and near infrared (NIR) data from the $u$-band to the WISE W2 band. The astrometric surveys within the {\it Gaia} mission significantly contributed to the determination of the extra-galactic nature of several quasar candidates, which are expected to show a parallax consistent with zero \citep[see][and references therein for further information on the selection principles]{Heintz2018b, Geier2019}. 
The present work is based on the identification of three objects along the slit of the OSIRIS instrument at the Gran Telescopio Canarias (GTC), used for the spectroscopic follow-up observations of quasar candidates  \citep{Geier2019}. The primary source, identified as GQ\,1114+1549A in the present paper, is a confirmed quasar candidate located at RA = 11:14:34.26, Dec = +15:49:44.80 (J2000.0) \citep{Heintz2018b}. The second source, hereby identified as GQ\,1114+1549B, had not been included by any previous survey and its quasar nature was confirmed in the present work. The spectrum of that object yielded a redshift, compatible with that of GQ\,1114+1549A, while having significantly different $g$, $r$ and $i$-band magnitudes compared to its companion. Ultimately, the spectrum of the third source was found to be reliably compatible with that of a foreground star.

The present paper aims at presenting the serendipitous discovery of GQ\,1114+1549B and complementing it with archival and follow-up observations, analysed with the aim of characterising the nature of the quasar pair. The small angular separation of this pair of quasars warrants further analysis, particularly with the objective 
of establishing whether the pair constitutes a gravitationally bound system. The present document is divided into three sections. Section \ref{sec:data} covers the observations conducted on the quasar pair, section \ref{sec:results} describes the main results, while section \ref{Sect:diskussion} concludes discussing the importance and implications of our discovery for future binary quasar studies.
We assume a $\Lambda$CDM cosmological model with $H_0=67.4$ km s$^{-1}$Mpc$^{-1}$, $\Omega_M=0.32$ and $\Omega_{\Lambda}=0.68$ \citep{Planck2018}.

\section{Observations} \label{sec:data}

The first spectroscopic observations of the quasar pair were obtained with OSIRIS instrument at the GTC. The candidate quasar GQ\,1114+1549A, selected according to optical data, NIR photometry and astrometry, was observed on December 4, 2018, when two 400 sec integrations were obtained. Using the Grism R1000B and a 1.23 arcsec slit, the observation yielded a resolution of $\mathcal{R}=500$, with a spectral range of 3750--7800 \AA. The observing conditions were a seeing of about 1.1 arcsec, and spectroscopic sky transparency, and dark time. The target was observed at an average airmass of 1.43, at parallactic slit angle.

While analysing the GTC spectrum of GQ\,1114+1549A, the authors found evidence for a second (adjacent) quasar on the slit. Follow-up observations of the quasar pair were obtained on July 2 and on July 5 2019 with the Nordic Optical Telescope (NOT). The total integration time was 2840 sec, with the aim of confirming the presence of a companion quasar in the field, aligning the slit with both quasars. Here, we used the low-resolution 
spectrograph AlFOSC and a grism covering the spectral range 3800--9000 \AA \ with a 1.3 
arcsec wide slit providing a resolution of $\mathcal{R}\approx280$. The pair was observed in evening twilight,
as it was already setting, with the position angle being 124$^\mathrm{o}$ East of North.
As the airmass was high (1.5--1.8) the observation suffered from significant differential slitloss, but the 
spectral region from 4500--9000 \AA \ was well covered.

The spectroscopic data from all the observations were reduced using standard procedures in 
IRAF\,\footnote{IRAF is distributed by the National Optical Astronomy
Observatory, which is operated by the Association of Universities for
Research in Astronomy (AURA) under a cooperative agreement with the
National Science Foundation.}. The GTC spectra were flux calibrated using 
observations of the spectro-photometric standard star Feige 110 observed on 
the same night. The NOT spectra were flux calibrated using a sensitivity function from an earlier night. 


To reject cosmic rays we used La\_cosmic \citep{vanDokkum01}.
We corrected the spectra for Galactic extinction using the extinction maps of \cite{Schlegel98}. To improve the absolute flux calibration we scaled the spectra to the $r$-band 
photometry from SDSS \citep{Alam15}.

\section{Results} \label{sec:results}

\begin{figure}
\epsscale{.95}
\plotone{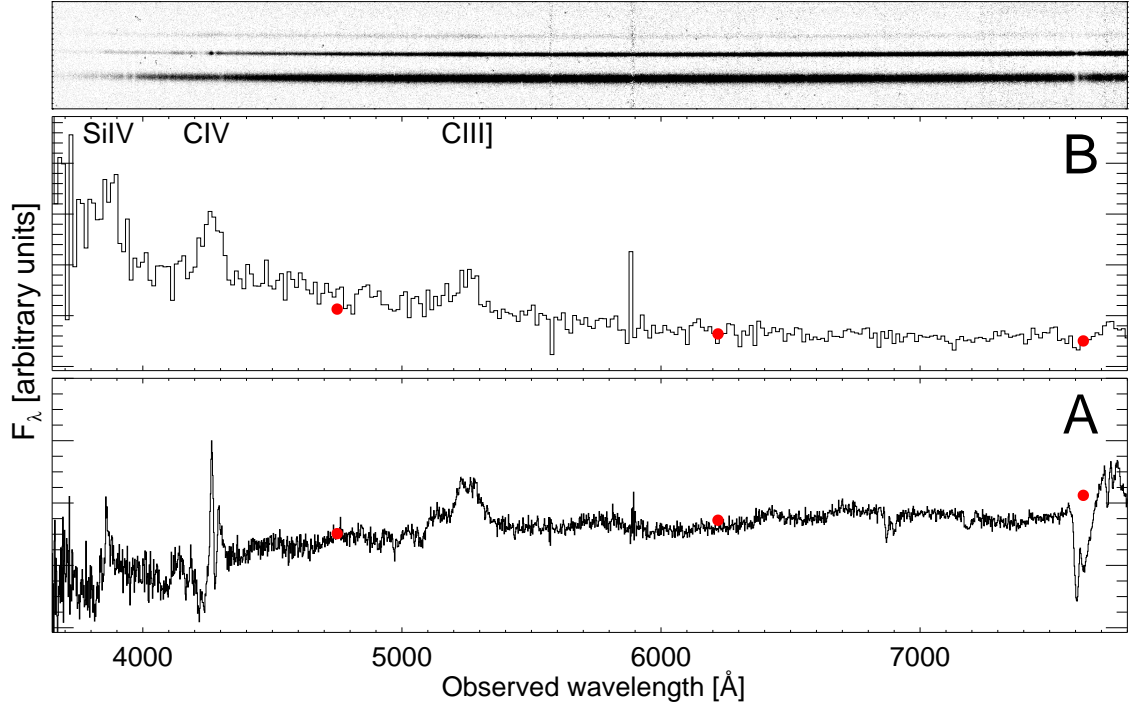}
\caption{Here we show the GTC spectra in which the \ion{Si}{4}, \ion{C}{4}, and \ion{C}{3}] lines are well covered. The top panel shows the 2-dimensional spectra showing three objects on the slit. The lowest trace is the bright star located west of GQ\,1114+1549A (see Fig.~\ref{fig:findchart}, the central trace is the primary quasar target (GQ\,1114+1549A) and the upper trace is the serendipitously discovered quasar (GQ\,1114+1549B). The two bottom panels show the quasar spectra, A (below) and B (above). The B-spectrum is suppressed by a about a factor of 10 (i.e., we get only 10\% of the flux in the slit compared to the A-spectrum). The B-spectrum is binned by a factor of seven for better visibility. Over-plotted is also the $g$, $r$, and $i$-band photometry from SDSS.}
\label{fig:2dspectrum}
\end{figure}

\begin{figure}
\epsscale{.80}
\plotone{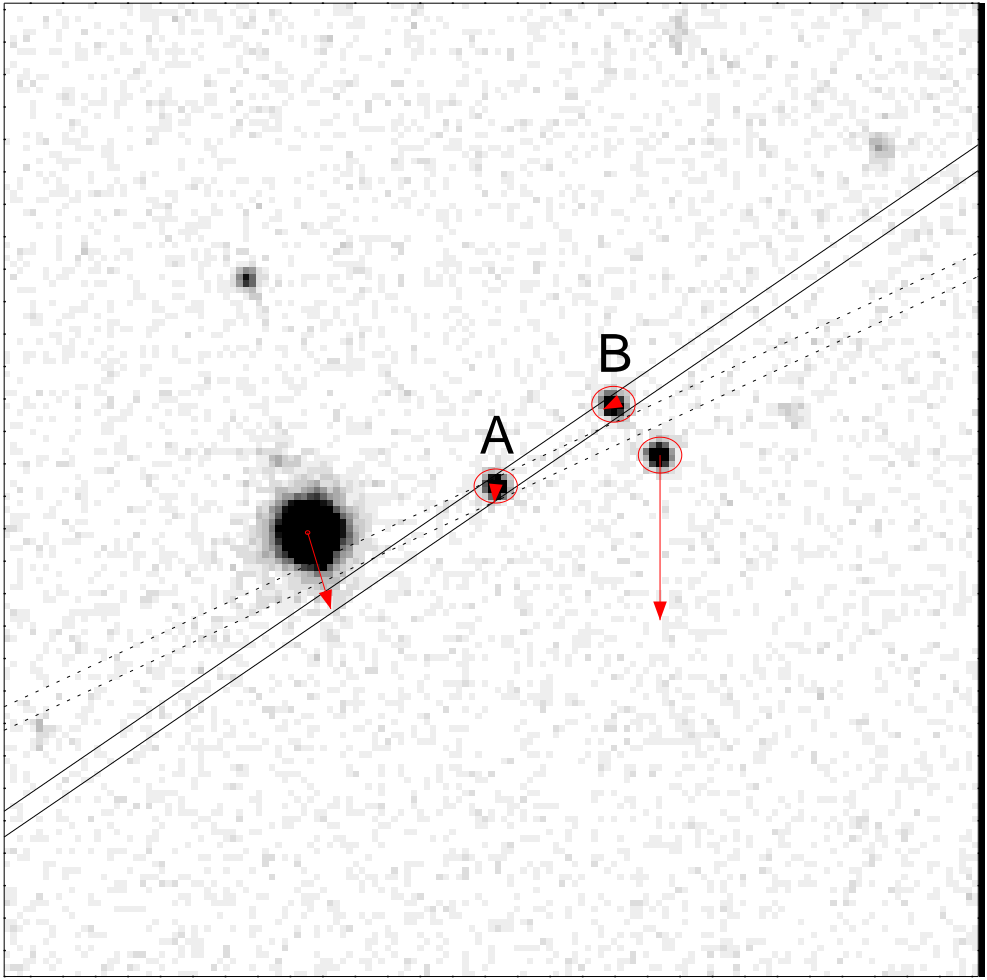}
\caption{A 1$\times$1 arcmin$^2$ field around the two sources (marked A and B)
as imaged in the $r$-band by SDSS DR12. North is up and East is to the left. We have over-plotted information on the proper motion from Gaia DR2 \citep{Gaia2018} with red arrows and red error ellipses showing the 2-$\sigma$ uncertainty on the proper motion.
The quasars both have proper motions consistent with 0, whereas the two other objects are moving significantly and are hence unrelated.
We also plot a schematic view of the slit during the GTC observation (dotted lines), which was centred on source A and aligned with the parallactic angle, oriented at 115$^\mathrm{o}$ East of North (EoN) at the time of the observation.
The position angle between the two objects is 124$^\mathrm{o}$ EoN, and this was the slit angle used during the NOT observation (full drawn lines).}
\label{fig:findchart}
\end{figure}

In Fig.~\ref{fig:2dspectrum} we show the discovery spectroscopy from GTC. The top panel shows the 2-dimensional spectra with the trace of GQ\,1114+1549A in the middle. Below that there is the trace of a bright star (see Fig.~\ref{fig:findchart}). 
Above the trace of GQ\,1114+1549A there is the weak trace of another source. In this trace, we noted the presence of possible broad emission lines indicative of this source being also a quasar. The two lower panels in Fig.~\ref{fig:2dspectrum} show the 
extracted 1-dimensional spectra of GQ\,1114+1549A and this possible second quasar on the slit (referred to as GQ\,1114+1549B).
As the trace is very weak we had to bin the spectrum significantly to properly bring the emission lines out of the noise for visibility.

In Figure~\ref{fig:findchart} we show a 1$\times$1 arcmin$^2$ field around the two sources
(marked with A and B) as imaged in the $r$-band by SDSS in DR12 \citep{Alam15}. We also indicate the slit orientations. From this image it is clear that the slit position covering GQ\,1114+1549A at the parallactic angle grazes another point source with similar brightness west of GQ\,1114+1549A, which must be GQ\,1114+1549B.

The projected angular separation between the two objects is 
$8.76\pm0.11~\mathrm{arcsec}$ as measured from the SDSS $r$-band image\footnote{http://skyserver.sdss.org/dr12/en/tools/explore/Summary.aspx?}.

\begin{figure}
\epsscale{.95}
\plotone{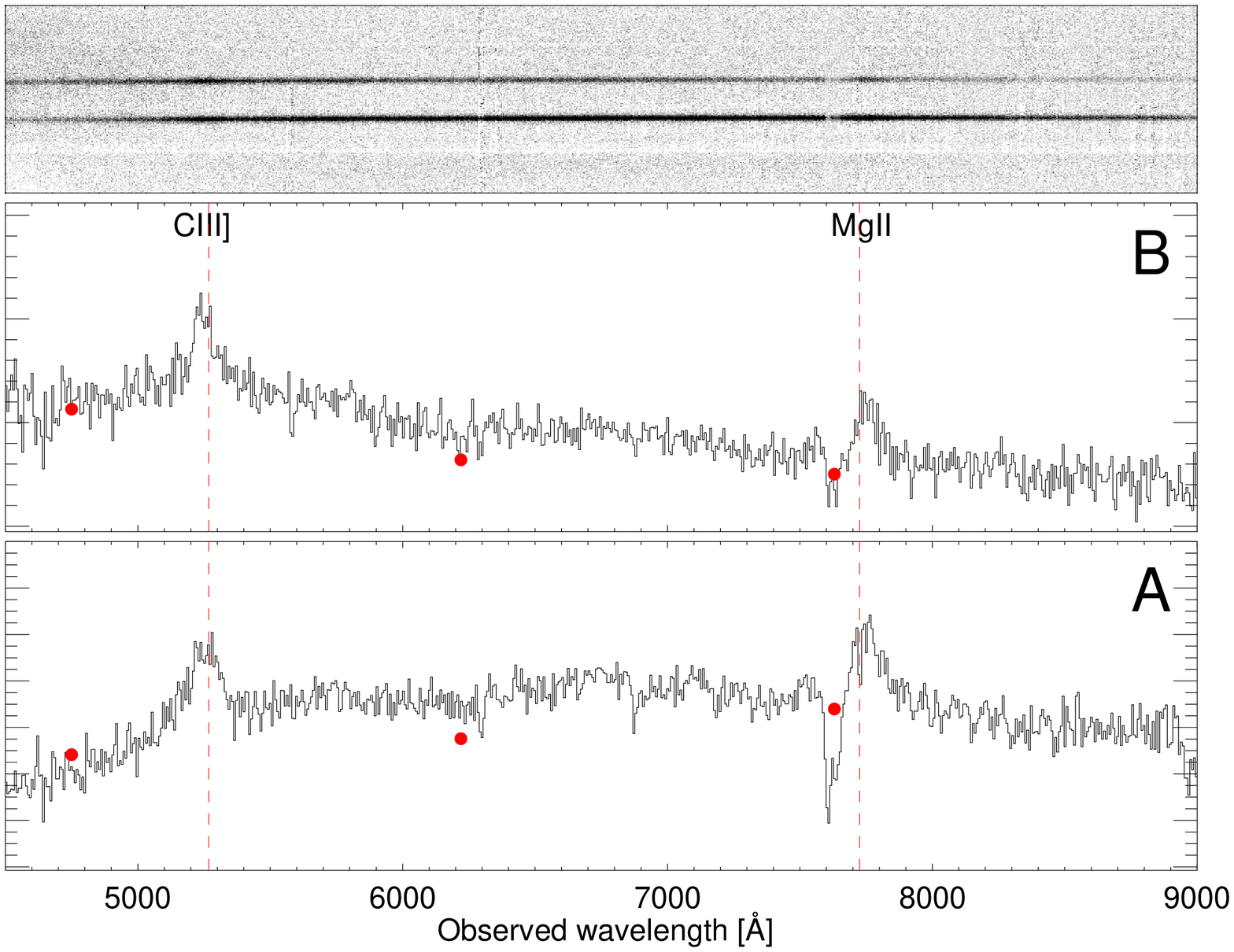}
\caption{Here are shown the spectra from the NOT obtained using a slit properly aligned with both quasars. The top panel illustrated the 2-dimensional spectrum with the traces of both quasars separated by 8.76 arcsec. The two bottom panels show the 1-dimensional spectra covering the region from \ion{C}{3}] to \ion{Mg}{2} (marked with dashed, red lines). The spectra are not corrected for telluric absorption. The red dots show the $g$, $r$ and $i$-band photometry from SDSS.}
\label{fig:2dspectrum_not}
\end{figure}

The purpose of the NOT spectroscopy was to obtain a spectrum with both GQ\,1114+1549A and GQ\,1114+1549B properly aligned.
As the source was already setting in evening twilight the observation was difficult, but the authors still managed to capture a useful spectrum, as shown in Fig.~\ref{fig:2dspectrum_not}. Here we again show both the 2-dimensional spectrum (top) and the extracted 1-dimensional spectra of both quasars. It is clear that both objects are quasars at a very similar redshift of 1.76$\pm0.01$.

By accounting for both GTC and NOT data, there is no doubt that a new physical binary quasar pair has been serendipitously discovered.
Remarkably, this is the second time that by chance we discover a quasar pair using a random slit angle \citep[for the first discovery see][]{Heintz2016}.

\section{Discussion and Conclusions}
\label{Sect:diskussion}

For two quasars at the same redshift at such a low projected separation the first question to answer is if this is a lensed source or a physical quasar pair. An overview of more than 200 known lensed quasars can be found in \citet{GaiaPairs2018}, but unfortunately the separations are not included in the list. Currently, a handful of lensed systems are known
with separation larger than 10 arcsec \citep{Inada2003,Inada2006,Dahle2013,Shu2018} so a separation of 8.76 arcsec does not exclude that the pair under study here is a lensed system. 

The redshifts of the two quasars are in our case also consistent within the errors. However, the spectra and spectral energy distributions, as mapped out by the photometry (see Table~\ref{tab:mag}), is remarkably different with one of the pair members (A) being strongly reddened and the other not (B). As an example, source A has a colour excess in the $r-z$ bands \citep[which is found to be a good tracer of reddening due to dust in high-$z$ quasars;][]{Heintz2018} of $r-z = 0.73\pm 0.01$\,mag, which is at the 95th percentile of the full sample of quasars from the SDSS data release 14 \citep{Paris18}. Source B with $r-z = 0.11\pm 0.01$\,mag falls well within one standard derivation from the mean compared to the same sample. Therefore, the system is most likely a physical pair of quasars with a projected proper distance of 75 kpc in the assumed cosmology.

\begin{table*}[!htbp]
\begin{tabular*}{1.0\textwidth}{@{\extracolsep{\fill}}l c c c c c c c c c }
    \noalign{\smallskip} \hline \hline \noalign{\smallskip}
    \emph{Object} & \emph{u} & \emph{g} & \emph{r} & \emph{i} & \emph{z} & \emph{Y} & \emph{J} & \emph{H} & \emph{K$_s$}\\
    & [mag] & [mag] & [mag] & [mag] & [mag] & [mag] & [mag] & [mag] & [mag]\\
    \hline
    A & 21.58 & 20.51 & 19.78 & 19.11 & 18.99 & 18.05 & 17.60 & 17.22 & 16.39 \\

    B & 19.99 & 19.83 & 19.86 & 19.68 & 19.75 & 18.96 & 18.75 & 18.24 & 18.48 \\
    \noalign{\smallskip} \hline \noalign{\smallskip}
\end{tabular*}
\caption{The optical and near-infrared magnitudes of object A and B (all on the AB magnitude system) from the SDSS and UKIDSS catalogues \citep{Warren2007,Alam15}.}
\label{tab:mag}
\end{table*}

Previously, \citet{Hennawi06,Hennawi2010} and \citet{Findlay2018} have carried out an extensive search for binary QSO systems, using the Sloan Digital Sky Survey \citep[SDSS;][]{York00} 
and the 2QZ QSO catalogues. 
The number of known physical quasar pairs includes many hundreds, but most of these have substantially larger separations than 10 arcsec \citep{Findlay2018}. Only about 30 pairs have separations smaller than 10 arcsec in the SDSS footprint where there are about half a million confirmed 
quasars \citep{Paris2018,Findlay2018}.
It is known that the number of quasar pairs with small separation (here $\lesssim$30 arcsec) is significantly underestimated in catalogues based
only on the SDSS/BOSS spectroscopic sample, primarily due to fibre collisions. \citet{Findlay2018} try to decrease this bias by searching for objects close to known quasars with quasar-like colours in the optical. However, significantly reddened quasars like
GQ\,1114+1549A are typically missed in quasar searches based only on optical colours. Indeed, neither GQ\,1114+1549A nor GQ\,1114+1549B are classified as quasars in SDSS. It is one of the goals of our selection, (which led to the discovery of GQ\,1114+1549A), to avoid the colour bias inherent in the SDSS quasar selection \citep[and to a lesser extend in the BOSS quasar selection, see also the discussion in][]{Krogager19}. Our previous serendipitous 
discovery of a quasar pair in \citet{Heintz2016} was also a system of two quasars with very different
colours, in that case, however, also with different redshifts. It is therefore possible that the number of quasar pairs at small separation is significantly 
underestimated. A new study based only on a selection of point sources without significant proper motion based on {\it Gaia} astrometric data next to known quasars is be a relatively easy way to explore this possibility. As seen in Fig.~\ref{fig:findchart}, astrometry is a very clean way of rejecting 
stars in searches for quasars \citep[see also][]{Heintz2018b}.

%
%

\section*{Acknowledgments}
\acknowledgments

The majority of the present work was carried out during a summer school hosted at the Astronomical Institute 
of the Slovak Academy of Sciences (Tatransk{\'a} Lomnica, Slovakia). Over the course of group projects, the first five authors identified the binary quasar, under the supervision of the 7\textsuperscript{th} author. We would like to
thank the organisers of the summer school, as well as acknowledging the support received from the ERASMUS+ grant number 2017-1-CZ01-KA203-035562, the European Union's Horizon 2020 research and the innovation programme, under grant agreement No 730890 (OPTICON).

We further thank A. Angeletti, J. X. Prochaska and L. Delchambre for helpful discussions. The presented work was made possible by observations conducted with the Gran Telescopio Canarias (GTC) and with the Nordic Optical Telescope (NOT), installed in the Spanish Observatorio del Roque de los Muchachos of the Instituto de Astrofísica de Canarias, in the island of La Palma. Our data analysis is partly based on data obtained through the UKIRT Infrared Deep Sky Survey. The Cosmic Dawn Center is funded by the DNRF. K. E. Heintz also acknowledges the support by a Project Grant (162948--051) from The Icelandic Research Fund. J. P. U. Fynbo thanks the Carlsberg Foundation for their support.

\bibliographystyle{aasjournal}
\bibliography{ms}

\end{document}